\documentclass[pra,final,twocolumn,showpacs]{revtex4-1}% Physical Review B
\usepackage{graphicx,amsmath,amsfonts,wasysym,color}
\usepackage{hyperref}
\begin{document}

%\preprint{APS/123-QED}

\title{Laser cooling of externally produced Mg ions in a Penning trap for sympathetic cooling of highly charged ions}% Force line breaks with \\

\author{Z. Andelkovic*}
\author{R. Cazan*}%
\author{W. N\"ortersh\"auser}%
\affiliation{Institut f\"ur Kernchemie, Universit\"at  Mainz, 55099 Mainz, Germany \\*These authors contributed equally to this work.}
\author{S. Bharadia, D.M. Segal, R.C. Thompson}%
\affiliation{Department of Physics, Imperial College London, SW7 2AZ London, United Kingdom}
\author{R. J\"ohren, J. Vollbrecht, V. Hannen}%
\affiliation{Institut f\"ur Kernphysik, Universit\"at M\"unster, 48149 M\"unster, Germany}
\author{M. Vogel}%
\affiliation{Institut f\"ur Angewandte Physik, Technische Universit\"at Darmstadt, 64289 Darmstadt, Germany}

\date{\today}% It is always \today, today,

\begin{abstract}
We have performed laser cooling of Mg$^+$ ions confined in a Penning trap. The externally produced ions were captured in flight, stored and laser cooled. Laser-induced fluorescence was observed perpendicular to the cooling laser axis. Optical detection down to the single ion level together with electronic detection of the ion oscillations inside the Penning trap have been used to acquire information on the ion storage time, ion number and ion temperature. Evidence for formation of ion crystals has been observed. These investigations are an important prerequisite for sympathetic cooling of simultaneously stored highly-charged ions and precision laser spectroscopy of forbidden transitions in these.
\end{abstract}

\maketitle

\section{Introduction}

Laser spectroscopy of optical transitions in highly charged ions (HCIs) is a subject of considerable interest as it provides access to relativistic effects in few-electron systems and can be used to test bound-state QED in the extremely strong electric and magnetic fields in the vicinity of the ionic nucleus \cite{Bei00,vogpr}. Experimentally, such magnetic dipole (M1) transitions in mid-$Z$ HCIs have first been studied in electron-beam ion traps (EBITs)\ by laser excitation and fluorescence detection \cite{maeckel2011}, yielding a relative accuracy of a few ppm for the  determination of the wavelength. Direct laser spectroscopy of heavy (high-$Z$) HCIs has so far only been performed at the experimental storage ring ESR on hydrogen-like bismuth $^{209}$Bi$^{82+}$ \cite{Kla94} and lead $^{207}$Pb$^{81+}$ \cite{See98}. In both cases, the transition between the ground state hyperfine levels was induced by pulsed lasers and resonance fluorescence was recorded. These investigations have been extended to the ground-state hyperfine transition in lithium-like bismuth $^{209}$Bi$^{80+}$, which has recently been observed in the experimental storage ring (ESR) \cite{Noe12}. This measurement in combination with the measurement on hydrogen-like bismuth will allow the first determination of the so-called 'specific difference' between the hyperfine splittings $\Delta E(1s,2s)$ as suggested by Shabaev and co-workers \cite{Sha01}. The first observation of the transition in $^{209}$Bi$^{80+}$ is an important step, but it will not provide sufficient accuracy for a high-precision determination of the QED effects in the specific difference, since the wavelength determination for both transitions (H-like and Li-like) is still limited in accuracy due to the large Doppler width and the uncertainty of additional Doppler shifts caused by the relativistic ion motion in the storage ring. This will be considerably improved once high-$Z$ highly charged ions are available at rest in a clean environment allowing for high-accuracy laser spectroscopy. To this end, the SpecTrap experiment has been designed \cite{vogel,zoran}. It is part of the highly charged heavy ion trap (HITRAP) project \cite{kluge} at the GSI Helmholtzzentrum Darmstadt, which will provide HCIs up to U$^{91+}$ at low energies suitable for capture into a Penning trap.

The precision achieved in the laser spectroscopy of trapped ions crucially depends on the width of the optical transition of interest and the mechanisms that lead to additional broadening, e.g.\ Doppler broadening. The study of forbidden transitions with high accuracy requires the elimination of Doppler broadening. This can be achieved by first-order Doppler-free techniques like two-photon transitions or by the trapping and cooling of atoms or ions. There is a variety of corresponding methods for the cooling of the ion motion, for a detailed overview see e.g.\ \cite{werth}. The evaporative cooling of HCIs in an EBIT has been used for the laser spectroscopy of Ar$^{13+}$ \cite{maeckel2011} and recently in a Penning trap on HCIs that were produced in an EBIT and then transported and re-trapped in a Penning trap \cite{hobein2011}. At SpecTrap we make use of resistive cooling \cite{win75,dan} and laser cooling \cite{itano,leib,esch,ric0}. The former is a very effective cooling mechanism for HCIs, while the latter is most effective for ions with a level scheme suitable for laser cooling such as Be$^{+}$ or Mg$^{+}$. Laser-cooled ions can then be used for sympathetic cooling \cite{werth} of simultaneously trapped HCIs. Such experiments have so far been performed with Be$^{+}$ in a Penning trap \cite{Gru05} and are foreseen in a Paul trap \cite{maeckel2011}. Here, we present first studies with laser-cooled Mg$^{+}$ ions in the SpecTrap Penning trap. We have performed systematic measurements with externally produced Mg ions which have been captured in flight and stored. the observation of laser-induced fluorescence (LIF) down to the single-ion level allows a determination of the ion storage time, ion number and ion temperature. Evidence for the formation of ion crystals has been observed. These measurements represent an initial characterization and optimization of the system as an important step towards the sympathetic cooling and precision laser spectroscopy of highly charged ions.

\section{Penning Trap Setup}
Penning traps are well-established tools for capture and confinement of externally produced ions. A static homogeneous magnetic field ensures radial confinement, while the electrode arrangement produces an electrostatic potential well which provides axial confinement of charged particles. Ions can thus be localized, which allows laser irradiation and fluorescence detection under well-controlled conditions. Stored ions can be motionally cooled to reduce the Doppler broadening of transition lines to well below the GHz level. The achievable storage time is fundamentally limited only by the residual gas pressure inside the trap, since collisions with gas particles may lead to ion loss. Typical storage times range from seconds to minutes, but also storage times of several months have been achieved \cite{haff03}. Hence, also slow transitions like magnetic dipole (M1) transitions can be investigated with high resolution and statistics. Such traps have been realized in numerous variations especially concerning their geometry, for details see \cite{ghosh,werth}. For the purposes of laser spectroscopy, trap geometries need to be chosen such that they allow both ions and light to enter and leave the trap suitably, as well as to provide the means for observing the fluorescence.

\begin{figure}[ht]
\begin{center}
%\begin{minipage}[c]{0.65\textwidth}
	\includegraphics[width=0.38\textwidth]{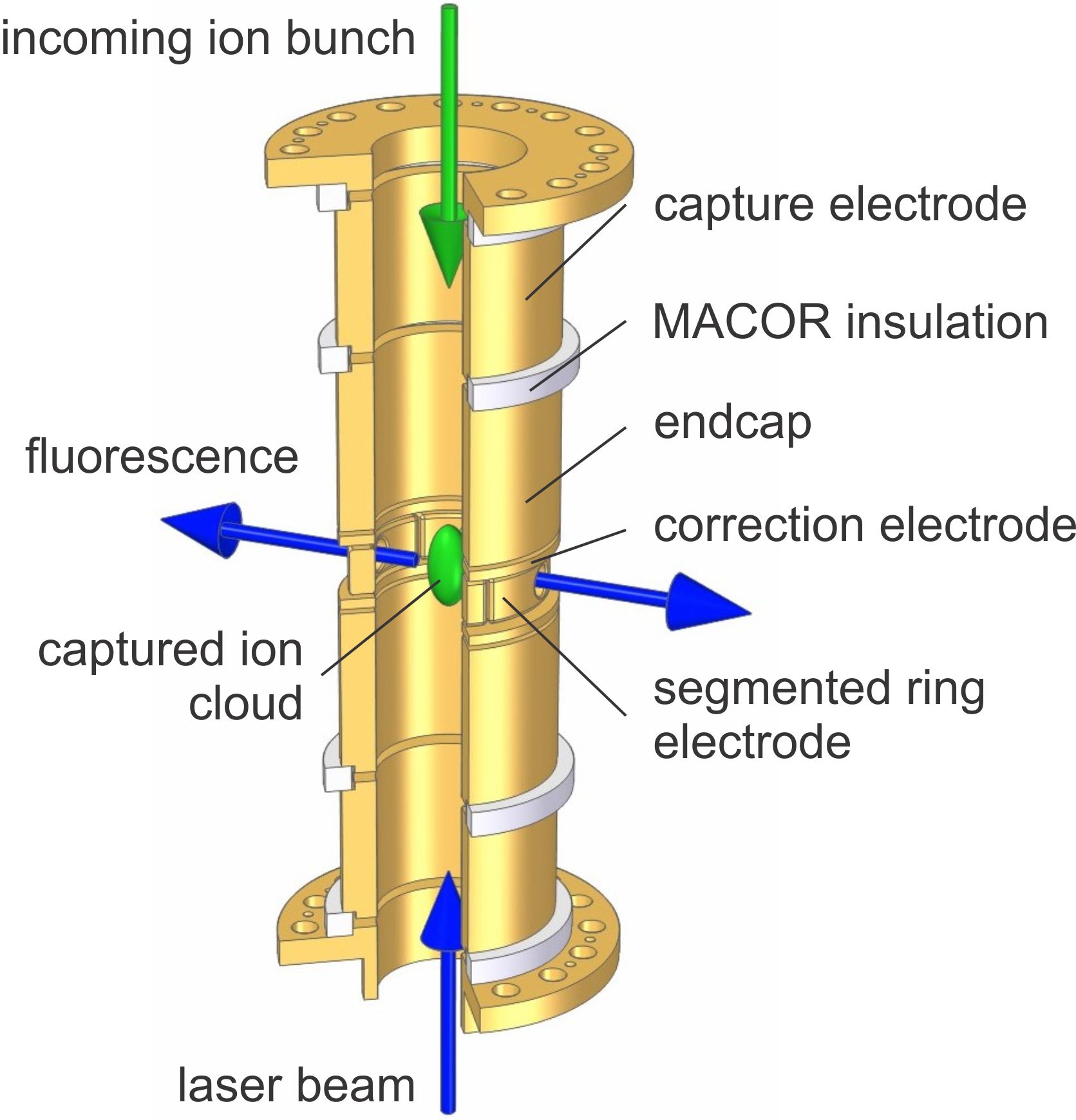}
%\end{minipage}
%\begin{minipage}[c]{0.3\textwidth}
	\caption{\small (Colour online) Overview of the Penning trap with the electrode stack and laser/ion beam orientation.}
	\label{fig:traplaser}
%\end{minipage}
\end{center}
\end{figure}

The SpecTrap experiment employs a five-pole cylindrical Penning trap with open endcaps \cite{gab89,brown86}, with an additional pair of capture electrodes, as described in detail in \cite{vogel,zoran}. The geometry is chosen such that the trap is orthogonal, i.e.\ the trapping potential depth is independent from the choice of correction voltages used to make the trapping potential harmonic close to the trap centre. The ion motion in such a trap has been discussed in detail in e.g.\ \cite{gab89, brown86, vogel}. The open endcaps and capture electrodes yield axial access to the trap from both sides. In our case, the ions enter from the top and the cooling laser from below, as shown in Fig.~\ref{fig:traplaser}. The capture of externally produced ions is achieved by fast switching of trap voltages. The ring electrode is radially split into four segments to allow the use of a rotating wall \cite{Bha12} for ion cloud compression and shaping. A central hole in each ring segment enables the detection of the stored ion fluorescence on radially positioned detectors outside the magnet vessel. The fluorescence light emerging out of the holes is collimated by plano-convex lenses. The geometrical light collection efficiency of this system is the main limiting factor of the total fluorescence detection efficiency.  Also, reflection and absorption in the lens and the vacuum windows as well as misalignments of the main optical axis reduce the signal. At the wavelength used for laser cooling of Mg${}^{+}$, the detection efficiency was measured to be about $\xi_{0}=3\cdot10^{-5}$.

The trap is installed in a vertical, cold-bore, superconducting magnet with Helmholtz configuration, such that direct optical access to the trap centre is possible through four radial ports in the horizontal plane. Before it was consigned to GSI, the magnet was used for a similar experiment (RETRAP at Lawrence Livermore National Laboratory), with a slightly different Penning trap configuration and radially cooled $\rm Be^+$ ions \cite{Wei98,Gru01,Gru05}. The magnetic field in the trap centre can be set to any value up to 6 T and provides a relative central homogeneity of $3\cdot10^{-5}$ over a region of 2.5~cm. A liquid helium cryostat is used for cooling both the superconducting solenoids and the trap with its attached electronics. The residual gas pressure in the vacuum system is monitored in the room temperature region at the bottom of the magnet vessel, and typically amounts to $5\cdot10^{-9}$~mbar during magnet operation. There is no direct separation between the trap and the insulation vacuum of the cryostat, so additional cryo\-pumping of the volume inside the trap is provided by the cold surfaces. Hence, the vacuum conditions inside the trap can be assumed to be much better than indicated by the gauge, as will be discussed below.

\begin{figure}[ht]
\begin{center}
%\begin{minipage}[c]{0.65\textwidth}
	\includegraphics[width=0.32\textwidth]{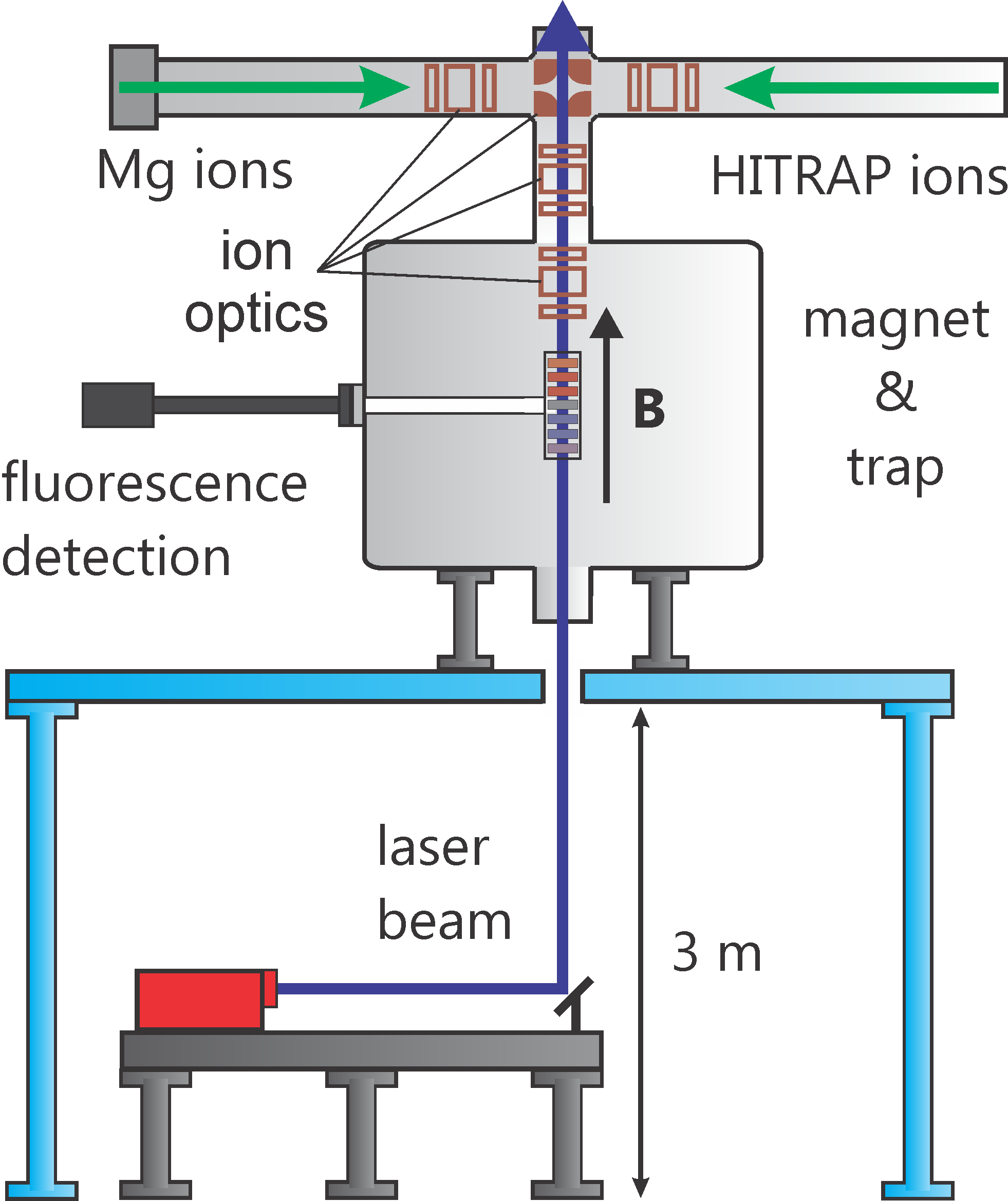}
%\end{minipage}
%\begin{minipage}[c]{0.3\textwidth}
	\caption{\small (Colour online) Overview of the SpecTrap experimental setup, including the ion beamline, the superconducting magnet and the laser laboratory (not true to scale).}
	\label{fig:trapsetup}
%\end{minipage}
\end{center}
\end{figure}

Laser beams are guided into the trap along the central vertical axis, from a laser laboratory located under the superconducting magnet setup, as shown in Fig.\ref{fig:trapsetup}. The fluorescence light is detected by a channel photo multiplier attached to the outside of the magnet vessel. It has a quantum efficiency of 18\% and a very low dark count of some 20 Hz. It is well suited for detection of UV light between 200 and 400 nm. Because of its sensitivity to the stray magnetic field it was mounted in a magnetically shielded housing about 1 metre away from the main magnet chamber, as depicted in Fig.~\ref{fig:trapsetup}.

To avoid excess heating of the cryostat and prepare the system for injection of externally produced HCI, Mg ions are produced by an off-line ion source. It consists of a directly heated tungsten crucible filled with grains of Mg metal. Mg atoms leaving the crucible are ionized inside a cup-formed grid by electrons emitted from a thoriated tungsten filament located outside the grid. The potential of the grid sets the energy of the produced ions. They are collimated with an einzel lens and enter a 90${}^{\circ}$ quadrupole deflector, which guides them into the vertical part of the beamline. The quadrupole geometry is chosen to allow injection of ions from both sides of the beamline and to have free access along the vertical axis for the laser beam. Two additional einzel lenses in the vertical beamline prepare the ion bunch for injection into the magnetic field and guide them into the trap. The second arm of the horizontal beamline will be connected to an EBIT and later to the HITRAP cooling trap in order to trap heavy HCI provided by the GSI accelerator facility.

\begin{figure}[ht]
\begin{center}
%\begin{minipage}[c]{0.61\textwidth}
	\includegraphics[width=0.48\textwidth]{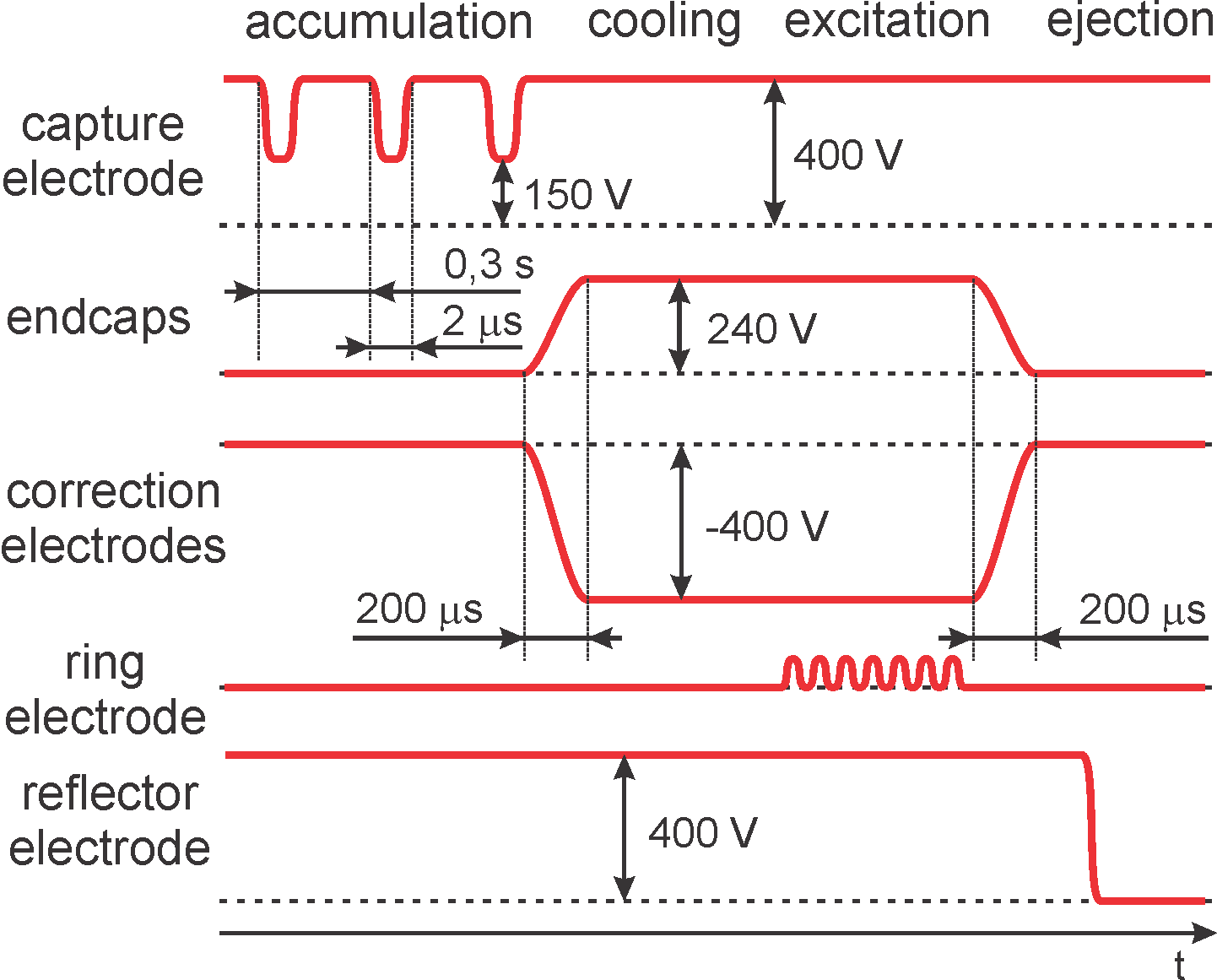}
%\end{minipage}
%\begin{minipage}[c]{0.37\textwidth}
	\caption{\small (Colour online) Overview of one trapping cycle. Voltage amplitudes on individual electrodes are plotted against time (both not true to scale). The dotted lines represent a reference level, typically ground.}
	\label{fig:trappingcycle}
%\end{minipage}
\end{center}
\end{figure}

Mg ions are produced in bunches of 1-2 $\mu$s length at a rate of a few Hz. They are transported towards the trap with a kinetic energy of 200 eV and dynamically captured into the Penning trap. One typical trapping cycle is illustrated in Fig.~\ref{fig:trappingcycle}. Initially, only the lower capture (reflector) electrode is permanently switched high (closed), while the upper capture electrode is switched between a confining potential and a value just below the ion transport energy, synchronized with the arrival time of the ion bunch. It has been experimentally observed that around 50~eV out of 200~eV axial energy are transferred into the radial motion during the ion injection into the magnetic field. That is sufficient for the accumulation of many ion bunches, with minimal losses of ions already stored during the reopening of the capture electrode. Typically 50-200 such accumulation cycles are repeated before permanently closing the capture electrode. The voltage on the endcaps and correction electrodes is then slowly (with respect to the ion motion) ramped up in order to compresses the ion cloud towards the trap centre. Afterwards, laser cooling by scanning the laser wavelength as well as electronic ion excitation and detection are performed.

\section{Laser cooling of ions in a Penning trap}
\label{sec:LCOIIAPT}
%Since HCI at low energies have a large cross section for electron capture which leads to their loss, a fast cooling method such as laser cooling is preferred.

\begin{figure}[ht]
\begin{center}
%\begin{minipage}[c]{0.61\textwidth}
	\includegraphics[width=0.45\textwidth]{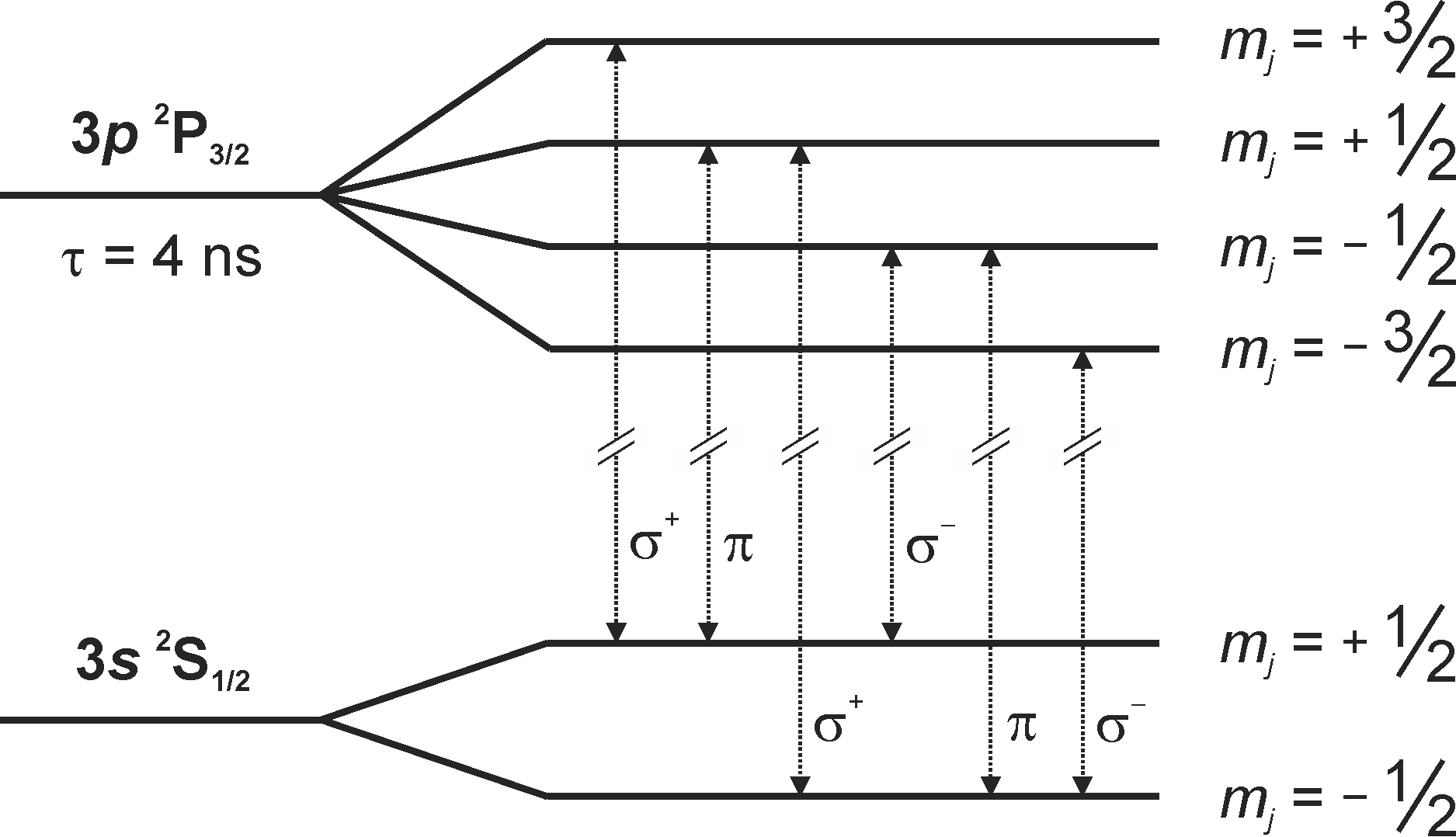}
%\end{minipage}
%\begin{minipage}[c]{0.37\textwidth}
	\caption{\small Zeeman splitting of the $3p^{2}\textrm{S}_{1/2}\rightarrow3p^{2}\textrm{P}_{3/2}$ ground state transition in $\rm Mg^+$. The necessary polarisation for driving the corresponding transition is also indicated.}
	\label{fig:mgzeeman}
%\end{minipage}
\end{center}
\end{figure}

A fast cooling method, such as laser cooling, is needed in order to rapidly decrease the energy of stored ions and reduce losses. As previously stated, direct laser cooling is limited to ions with a favourable level structure. Such ions can then be used for sympathetic cooling of other ions of interest which are simultaneously stored. A suitable species for laser cooling is the $^{24}$Mg$^+$ ion. It can be easily produced and the undisturbed ion provides a closed, ground-state, two-level $3s\ ^2\textrm{S}_{1/2}$ - $3p\ ^2\textrm{P}_{3/2}$ transition, with an excited state natural lifetime of only 4 ns. However, magnetic fields of several Tesla in the Penning trap lead to splitting of the Mg sublevels due to the Zeeman effect, as shown in Fig.~\ref{fig:mgzeeman}. The Zeeman slope coefficient for each sublevel is provided in Table~\ref{zeemanshift}. 

\begin{table}[ht]
	\centering
	\caption[The Zeeman shifts of $^2$S$_{1/2}$ and $^2$P$_{3/2}$ in $^{24}$Mg$^+$]{The Zeeman shifts of the $m_j$ levels in an external magnetic field for $^2$S$_{1/2}$ and $^2$P$_{3/2}$ in $^{24}$Mg$^+$.}
		\begin{tabular} {ccc}
			{\bf Level} & {\bf $m_j$} & {\bf {$\Delta E_{\rm{ZE}}/{hB}$(GHz/T)}} \\
			\hline
			$^2$S$_{1/2}$ & $-1/2$ & $-13.996$  \\
			$^2$S$_{1/2}$ & $+1/2$ & $+13.996$ \\			
			$^2$P$_{3/2}$ & $-3/2$ & $-27.992$  \\
			$^2$P$_{3/2}$ & $-1/2$ & $-9.331$ \\
			$^2$P$_{3/2}$ & $+1/2$ & $+9.331$  \\
			$^2$P$_{3/2}$ & $+3/2$ & $+27.992$  \\
			\end{tabular}
	\label{zeemanshift}
\end{table}

In this level scheme only the $\pm\frac{1}{2}\rightarrow \pm\frac{3}{2}$ (here and in further text the $m_j$ quantum numbers) Zeeman transitions remain closed systems. We have chosen the lowest transition $-\frac{1}{2}\rightarrow-\frac{3}{2}$ for the cooling process. It has a Zeeman shift of $-13.996$~GHz/T compared to the unperturbed $^2$S$_{1/2}\rightarrow^2$P$_{3/2}$ transition frequency of \linebreak $\nu_0=1~072~082.934$~GHz \cite{Mgmetr}. Considering the required polarization for the cooling laser and the injection direction of the ions, the laser is polarized $\sigma^-$ and sent along the trap axis.

A specific issue of the experiment is the relatively high kinetic energy of the captured ions, required for an efficient transport and a small ion bunch width. Frequency detuning of the laser corresponding to a typical transport energy of 200~eV together with a very fast adjustment of this detuning to match the dropping energy caused by the cooling would be a serious technical challenge. However, a much simpler approach can be employed at the expense of the cooling speed: since the axial speed of the injected ions varies between a maximum corresponding to 200~eV at the trap centre and zero at the turning points near the endcaps, the laser can be kept fixed at a frequency corresponding to a cold ion. It can then absorb a photon near each turning point \cite{Radu} which is still efficient provided that the laser intensity is sufficiently large. The condition is that the Rabi frequency for this transition is much larger than the axial frequency of the ion inside the trap, $\Omega \gg \omega_z$. It should be noted that only the axial ion motion will be directly cooled this way, since there is no cooling force acting on the ion cloud in the radial direction.

\begin{figure}[ht]
\begin{center}
	\includegraphics[width=0.43\textwidth]{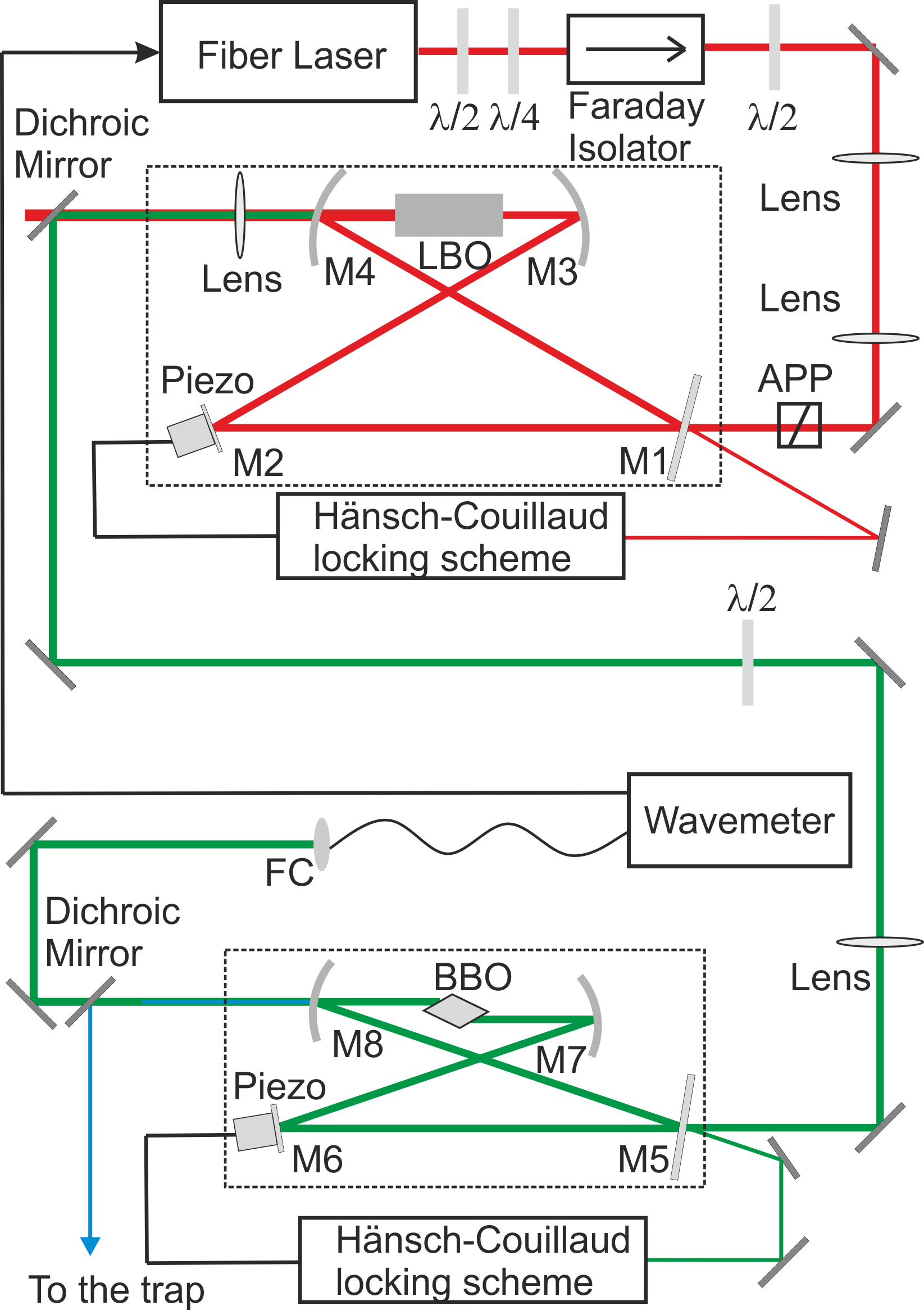}
	\caption{\small (Colour online) Setup of the laser system for cooling of Mg${}^{+}$: the laser beam from a fiber laser is frequency quadrupled using two non-linear crystals in bow-tie resonators with the length stabilized using the H\"ansch-Couillaud locking scheme. The wavelength is controlled using a High-Finesse wavemeter. ($\lambda/4$ and $\lambda/2$: quarter- and half-wave plate respectively, APP: anamorphic prism pair, M1-M8 cavity mirrors, FC: fiber coupler)}
	\label{fig:laser}
%\end{minipage}
\end{center}
\end{figure}

An all-solid-state laser system at the required wavelength of 279 nm has been set up as depicted in Fig.~\ref{fig:laser}. It has been described in detail in \cite{Caz10,Radu} and comprises a single-mode fiber laser at 1118~nm as well as two cavities for second harmonic generation to obtain frequency quadrupling. The laser is a Koheras Boostik fiber laser, specified to deliver 1.66~W maximum output power. Experimentally, a maximum of 1.2~W including the amplified spontaneous emission (ASE) was obtained.

Since the doubling efficiency for second harmonic generation (SHG) inside a non-linear crystal is proportional to the square of the fundamental power, bow-tie optical resonators were constructed to enhance the laser power inside the crystal. The first doubler uses non-critical phase-matching in a lithium triborate (LBO) non-linear crystal, which has a phase-matching temperature at 1118~nm of $\approx$~90$^\circ$C. The second doubler employs critical phase-matching of a beta barium borate (BBO) non-linear crystal, using the round, Gaussian output of the first doubler. Both resonators were designed using the ray transfer matrix analysis and computer simulations in order to obtain the optimal experimental parameters according to the Boyd-Kleinmann theory. The length of the doublers is actively stabilized using H\"ansch-Couillaud polarization-analysis locking \cite{HC80}.

For the first doubler, a maximum overall SHG efficiency of 33\% was obtained, providing 320~mW of laser power at 559.3~nm from 950~mW of the fundamental 1118.5~nm power. For the second doubler, a maximum of 16.7~mW at 279.6~nm was achieved using 210~mW of green power in front of the resonator, equivalent to an overall SHG efficiency of 8\%. All power levels of the harmonics were measured after appropriate filtering of the fundamental power leaking from the bow-tie resonators. In spite of several problems with the Koheras main fiber laser, which greatly affected the available pump power, a sufficient amount of about 2~mW UV laser power was available for the first trapping tests. The parameters of the frequency quadrupling system are summarized in Table~\ref{resonatorsummary}.
\begin{table}[t]
	\centering
	\caption[Summary of the most relevant parameters of the two resonators]{Summary of the most relevant parameters of the frequency quadrupling system.}
	\label{resonatorsummary}
		\begin{tabular} {ccc}
			{\bf Parameter} & {\bf $1^{st}$ Resonator} & {\bf $2^{nd}$ Resonator} \\
			\hline
			Crystal type & LBO & BBO  \\
			Phase matching & NCPM Type I & CPM Type I  \\
			Crystal length & 20~mm &	7.4 mm  \\
			Input wavelength $\lambda_0$ & 1118.54 nm & 559.27 nm \\
			Crystal cut $\theta$ & 90$^\circ$ & 44.4$^\circ$ \\
			Crystal cut $\phi$ & 0$^\circ$ & 0$^\circ$ \\
			Crystal surfaces & Dual AR & Brewster-cut\\
			Crystal temperature & 96$^\circ$C & 50$^\circ$C \\
			Total cavity length & 1158 mm & 504 mm\\
			Focusing mirrors $f$ & 70 mm & 50 mm\\
			Focusing arm length & 157.5 mm & 104.8 \\
			Full folding angle & 38$^\circ$ & 18.8$^\circ$ \\
			Enhancement factor $A$ & $\approx$60 & $\approx$70 \\
			Coupling efficiency  & $>$85\% & $>$85\% \\ 
			Input power ($\omega$) & 950 mW & 210 mW \\ 
			Output power ($2\omega$) & 320 mW & 16.7 mW \\
			Doubling efficiency & $\approx$33$\%$ & $\approx$8$\%$ \\		   			 
		\end{tabular} \\		
\end{table}

\section{Experimental Results}

\subsection{Ion storage time}
The storage time constant has been determined by monitoring the laser induced fluorescence of the trapped ions as a function of time. For this measurement, the laser has been tuned to a frequency $\approx$~200~MHz above the resonance frequency of the $-\frac{1}{2}\rightarrow -\frac{3}{2}$ transition in order to avoid strong laser cooling and the resulting fluctuation of the fluorescence due to the decreasing Doppler width of the transition. On the other hand, although a blue-detuned laser frequency leads to heating of the ion cloud, it is important to note that it is still red-detuned for the other Mg isotopes which might be present. Additionally, resistive cooling contributes to ion cooling with a time constant of around 100~s for $\rm Mg^{+}$ stored in SpecTrap \cite{AndThe}. Thus, an equilibrium between heating and cooling processes is established, resulting in a fluorescence signal proportional to the number of stored ions.

\begin{figure}[ht]
\begin{center}
%\begin{minipage}[c]{0.61\textwidth}
	\includegraphics[width=0.42\textwidth]{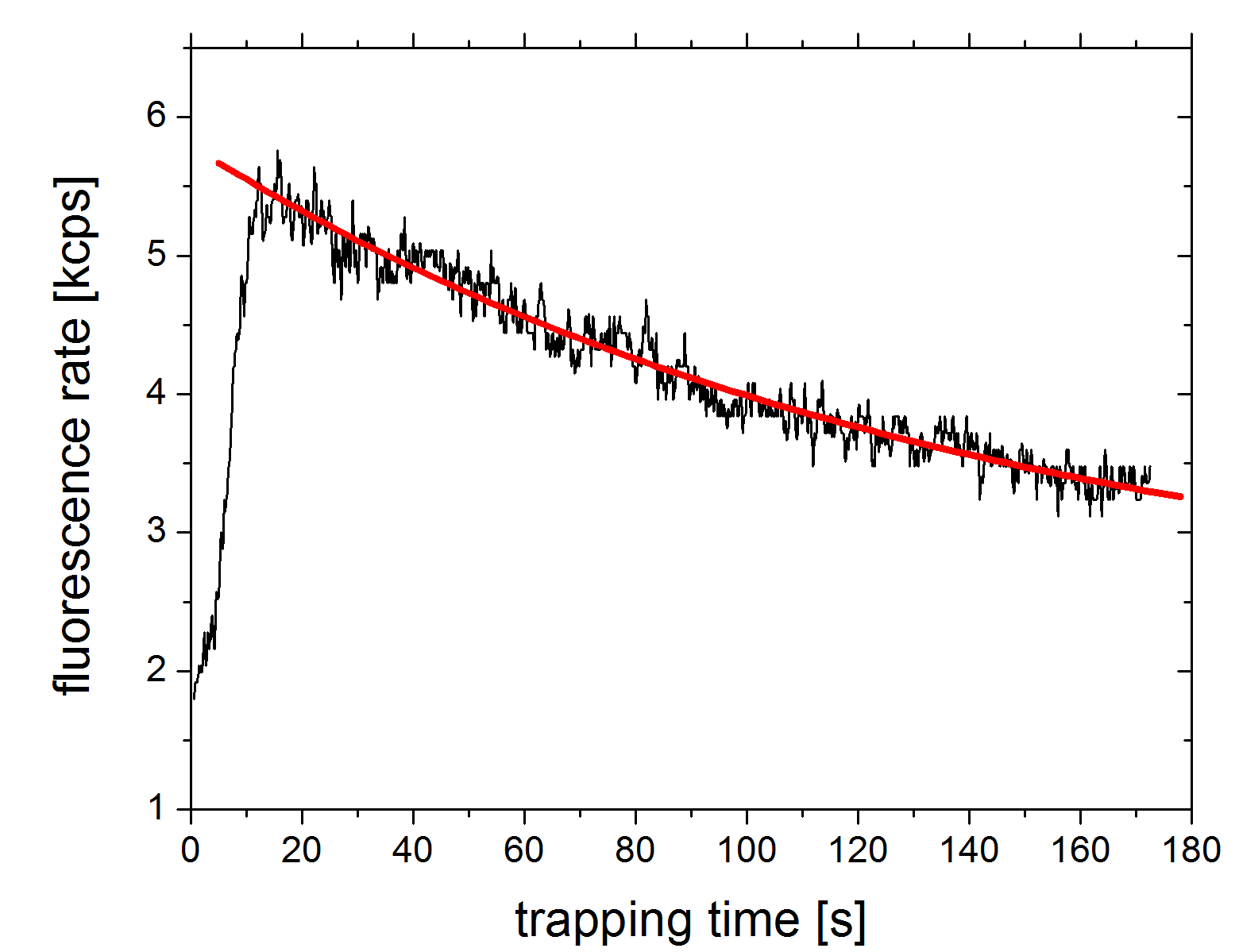}
%\end{minipage}
%\begin{minipage}[c]{0.37\textwidth}
	\caption{\small (Colour online) Storage time determination for Mg${}^+$ stored in SpecTrap. The $-\frac{1}{2}\rightarrow -\frac{3}{2}$ transition was continuously excited with a 200~MHz blue-detuned laser. The observed fluorescence rate is plotted as a function of time. At $t=0$ loading was complete and the endcaps were closed. The data was fitted with a single exponential function for points with $t\geq20$~s and the storage time constant of 137(5)~s was observed.}
	\label{fig:ionlifetime}
%\end{minipage}
\end{center}
\end{figure}

A typical trace resulting from this procedure is shown in Fig. \ref{fig:ionlifetime}. A storage time constant of about 140~s can be extracted from a single exponential fit to the data. Since in this measurement the ions were not cooled to sub-K temperature it can be regarded as a lower limit for the ion lifetime in the trap.

After improving the laser stability and the vacuum conditions, longer storage times of up to an hour have recently been observed for a cold cloud of $\rm Mg^+$ ions \cite{Mur11}. Further efforts in this direction are ongoing.

Unlike $\rm Mg^+$, the storage time of HCI can be significantly smaller and an estimate should be made. Ion loss is mainly attributed to charge exchange with residual gas particles. Although the cross sections for electron capture in ion-neutral collisions at very low energies are largely unknown, they can be estimated using the semi-empirical M\"uller-Salzborn formula \cite{Mue77}
\begin{equation}
	\sigma_S = 1.43\cdot 10^{-16} \cdot q^{1.17} \cdot I^{-2.76} \ \left[\textrm{m}^2\right]
\label{eq:schlachter}
\end{equation}
where $q$ is the charge state of the ion and $I$ is the ionization potential of the residual gas particle expressed in eV, which amounts to $I=15.44$~eV for H${}_2$ \cite{Her69} and $I=25.59$~eV for He. Partial pressures of all other typical residual gases are much smaller at the cryogenic temperature around the trap, and can be safely neglected. Alternatively, the cross section can also be estimated using the so-called classical barrier model \cite{Man86,Wei98} which brings similar results.

The rate $k_{\rm ec}$ of electron capture is then calculated by multiplying the cross-section $\sigma$ from Eq.~\eqref{eq:schlachter} by the neutral particle density $n$ and the relative velocity $v_r$ of the two colliding particles. The expected ion storage time is given through the reciprocal value of this rate
\begin{equation}
\tau = k_{\rm ec}^{-1} = \frac{1}{\sigma n v_r} =
	\frac{k_BT_r}{\sigma p}\left(\frac{3k_BT_i}{m_i}+\frac{3k_BT_r}{m_r}\right)^{-\frac{1}{2}}
\label{eq:ionlifetime}
\end{equation}
where $k_B$ is the Boltzmann constant, $p$ the pressure, $m_{r/i}$ and $T_{r/i}$ are the mass and temperature of the residual gas atoms and the ions, respectively. 

Assuming that the pressure inside the trap volume is not worse than $\approx 2 \cdot 10^{-11}$~mbar, which corresponds to $10^{-9}$~mbar measured in the 300~K region and scaled down to liquid helium temperature, the lifetime of around 160~s and 18~s is calculated using Eq.~\eqref{eq:ionlifetime} for charge states $q=13$ and $q=82$, respectively. These lifetime-estimates were calculated for ion temperatures around 1~K or less and rapidly decrease with increasing ion temperature, pointing towards the need for rapid ion cooling, such as sympathetic cooling with laser cooled $\rm Mg^+$.

\subsection{Laser cooling time and power}
The laser cooling time can be estimated by evaluating the cooling force exerted by the photons. Generally, for low energy ions, the scattering of the photons leads to a frictional force 
\begin{equation}
	F_{\rm{scatt}}=(\rm{photon~momentum})\times(\rm{scattering~rate})
	\label{fscatt0}
\end{equation}
which slows the ion down. It can be written as \cite{foot} 
\begin{equation}
	F_{\rm{scatt}}=\hbar k_L \,\frac{\Gamma}{2}\,\frac{I/I_{\rm{sat}}}{1+I/I_{\rm{sat}}+4\delta^2/\Gamma^2},
	\label{fscatt}
\end{equation}
where $\hbar k_L$ is the photon momentum, $\Gamma$ is the transition linewidth, $I$ is the intensity of the laser, $I_{\rm{sat}}$ the saturation intensity and $\delta$ the detuning of the laser frequency. This force is proportional to the laser intensity below the saturation value and it approaches its maximum value 
\begin{equation}
	F_{\rm{max}}=\hbar k_L \frac{\Gamma}{2}
	\label{fscattmax}
\end{equation}
for intensities $I \gg I_{\rm{sat}}$.

Here, the kinetic energy of the captured ions is typically 200~eV, and the laser frequency cannot be scanned fast enough to maintain the cooling condition. The laser is thus kept fixed at a small red-detuning and the scattering force in Eq.~(\ref{fscatt0}) can then be written as
\begin{equation}
	F_{\rm{scatt}}=\hbar k_L \omega_{\rm scatt},
	\label{fscattax}
\end{equation}
where $\omega_{scatt}$ is a scattering frequency inversely proportional to the ion velocity. This equation is a good approximation for an intensity close to or above the saturation intensity. Since under our experimental conditions the laser intensity was about 1/3 of the saturation intensity, the cooling force is reduced by a factor of $I/(I+I_{\rm{sat}})=1/4$. The deceleration can then be expressed as
\begin{equation}
	a(I_{\rm{sat}}/3)=\frac{F_{\rm{scatt}}(I_{\rm{sat}}/3)}{m}
	=\frac{1}{4}\frac{h\omega_{\rm scatt}}{\lambda m}
	\label{eq:decel}
\end{equation}
where $\lambda$ is the wavelength of the cooling laser and $m$ the mass of $^{24}$Mg$^+$. The stopping time can be calculated accordingly as
\begin{equation}
	t_0	= \int_{0}^{v_0}\frac{{\rm d}v}{a(I_{\rm{sat}}/3)}
			= \frac{4\lambda m}{h}\int_{0}^{v_0}\frac{{\rm d}v}{\omega_{\rm scatt}}
\end{equation}
where $v_0$ is the initial speed of the ions. If $\omega_{scatt}$ is approximated as the maximal scattering frequency reduced by the ratio between the natural linewidth $\Gamma$ and the Doppler width $\nu_D$ of the transition, in our case this stopping time approximation results in $t_0\approx 40$~s. 

The experimental cooling time was determined using the same measurement procedure as described in the previous section for recording the LIF signal shown in Fig.~\ref{fig:ionlifetime}. Initially, the ions have a large spatial oscillation amplitude between the endcaps while the fluorescence detection system is focused on a small volume at the centre of the trap. Hence, the emitted photons cannot be recorded and only the background signal is present. As the ions are cooled, they get localized in the centre of the trap and the fluorescence rate per ion rises as the Doppler-shifted transition matches the fixed red-detuned laser frequency for an increasing amount of ions. This results in the sharp rise in fluorescence observed in Fig.~\ref{fig:ionlifetime}, which appears about 10 seconds after raising the endcaps to the trapping potential. The measured value is of the same order of magnitude, but smaller than the predicted 40 s because of uneven ion velocity distribution and a finite probability for photon absorption also outside the natural linewidth of the transition.

\subsection{Single ion fluorescence}
\begin{figure}[ht]
\begin{center}
%\begin{minipage}[c]{0.61\textwidth}
	\includegraphics[width=0.23\textwidth]{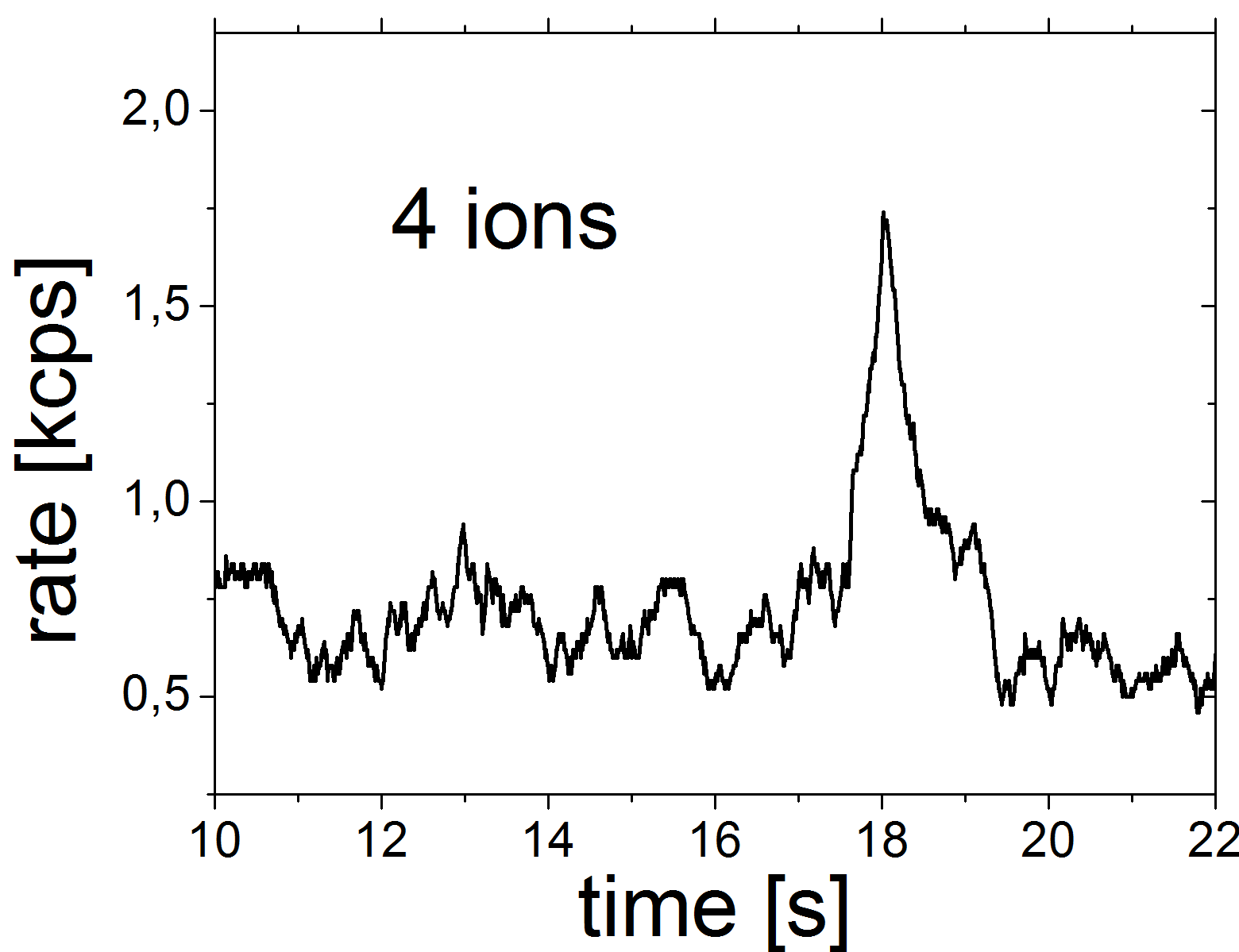}
	\includegraphics[width=0.23\textwidth]{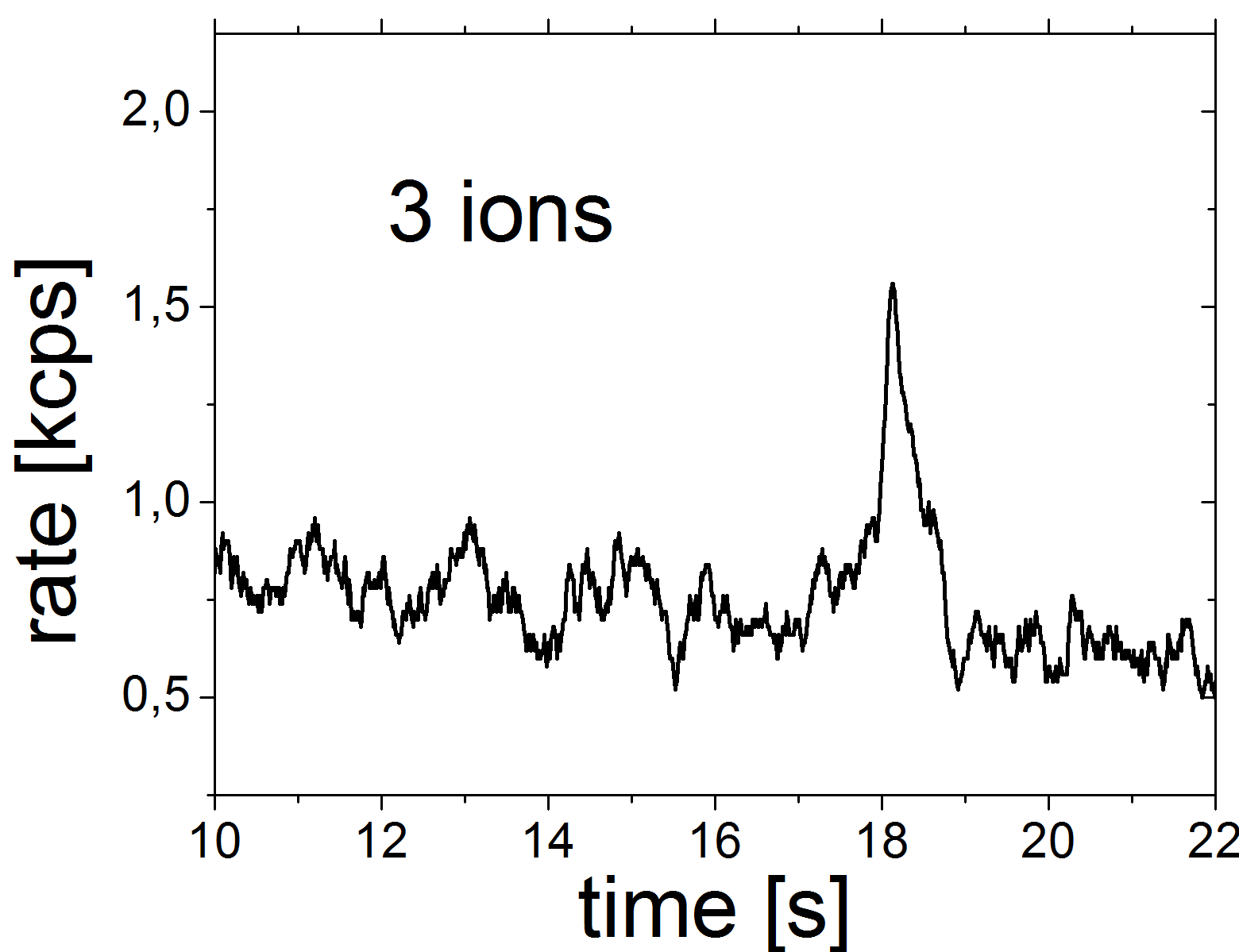}
	\includegraphics[width=0.23\textwidth]{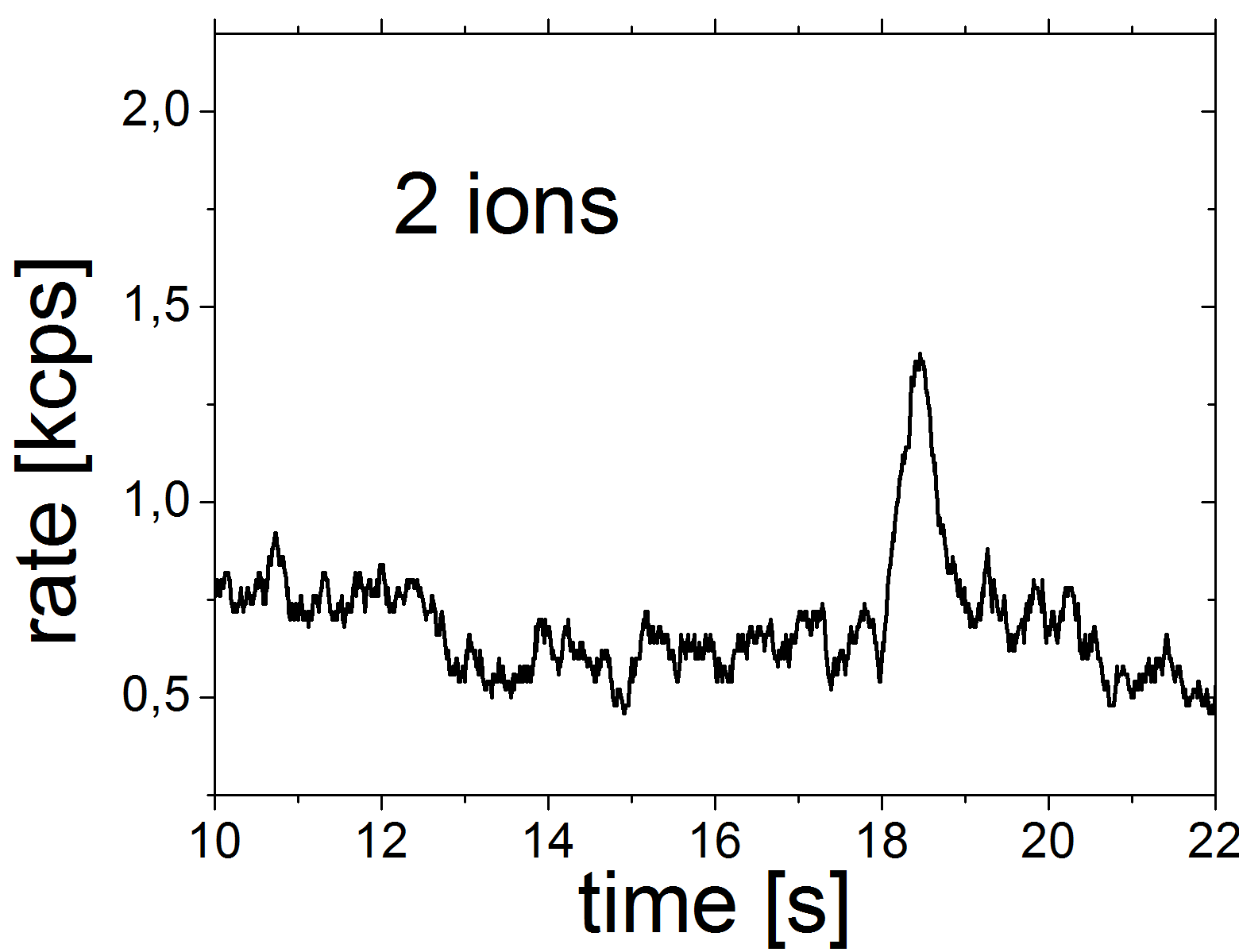}
	\includegraphics[width=0.23\textwidth]{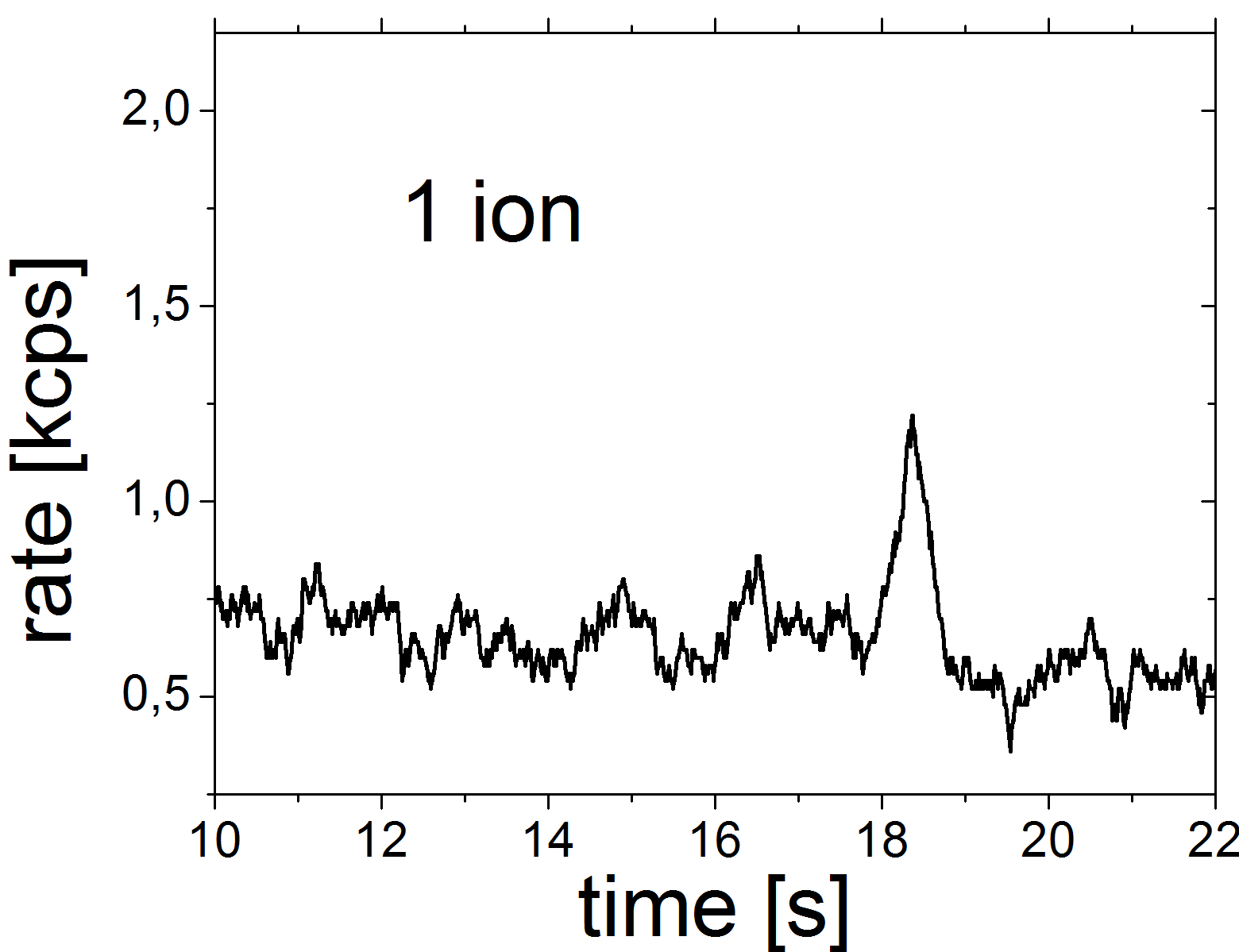}
%\end{minipage}
%\begin{minipage}[c]{0.37\textwidth}
	\caption[Single trapped ions]{\small Spectra with quantized fluorescence leaps of ions trapped in SpecTrap, associated with single trapped ions. The \textit{x}-axis is directly proportional to frequency with a factor of 100 MHz/s. The resonant frequency is reached after about 18~s.}
	\label{fig:singleions}
%\end{minipage}
\end{center}
\end{figure}

In contrast to trapping a large ion cloud, with the current experimental setup it was also possible to isolate and observe the fluorescence of a single trapped ion. The spectra shown in Fig.~\ref{fig:singleions} were recorded by scanning the laser from -1~GHz with 100~MHz/s across the resonance, using a 0.9~mW laser beam with a diameter of $\approx 1$~mm. Taken under identical conditions and trapping times, they show quantized changes of the laser induced fluorescence, associated with single trapped ions on top of a constant background signal.

\begin{figure}[ht]
\begin{center}
	\includegraphics[width=0.42\textwidth]{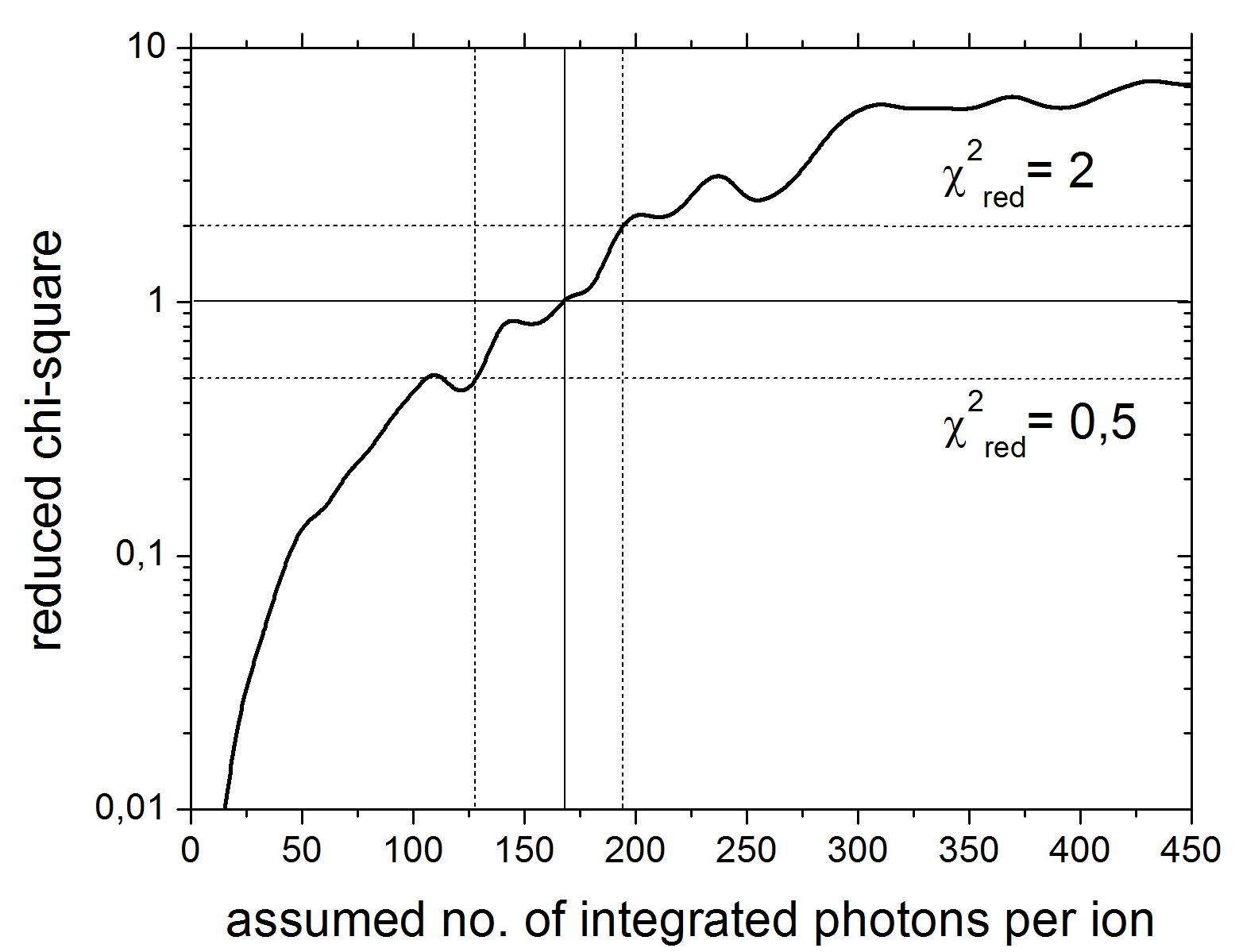}
	\caption[Number of fluorescence photons per trapped ion]{\small Distribution of the reduced chi-square for different assumed numbers of integrated photons per ion. The marked area around $\chi^2_{\rm red}=1$ represents the most probable value of roughly $170\pm 35$ photons.}
	\label{fig:chisquare}
\end{center}
\end{figure}

In order to verify that the single-ion regime was reached the following procedure was carried out: first an integrated number of detected photons over the resonance was determined for each of the recorded few-ion spectra. The error was treated as the statistical uncertainty with an added offset from the average deviation of the background. The number of photons per ion was then varied between 10 and 500 and compared to the recorded spectra, producing a deviation for each point. These deviations were used to calculate the reduced chi-square for each assumed number of photons per ion. The result of this procedure is plotted in Fig.~\ref{fig:chisquare}. The area around $\chi^2_{\rm red}=1$ corresponds to the most probable number of integrated photons per ion (roughly $170\pm 35$) which was used to determine the number of ions in the spectra shown in Fig.~\ref{fig:singleions}.
According to it, a single stored $\rm Mg^+$ ion in full resonance yields a rate of around $500$ fluorescence photons per second, on top of around $700$ background photons per second. The uncertainty of this value is dictated by the fluctuation of the background and the frequency uncertainty of the scanning laser frequency. Having that in mind it can be concluded that under the given conditions a single ion yields a fluorescence signal of $500 \pm 100$ photons/s.

In spite of the large uncertainty of around 20\%, this result shows that even a few trapped ions with a fast optical transition can be detected in SpecTrap via laser induced fluorescence. If the observed number of fluorescence photons is compared to the maximum expected number of photons from a non-fully saturated $\rm Mg^+$ ion, this yields a total detection efficiency of $\xi_{\textrm{tot}} = 5\cdot 10^{-6}$. Disregarding the quantum efficiency of the detector and considering only geometrical factors the detection efficiency amounts to about $\xi_{0} = 3\cdot10^{-5}$. Additionally, the expected number of photons per ion can be used to quantify a signal from an ion cloud and estimate the number of ions stored under the same conditions. A maximum of about 2000 ions were trapped and cooled using the current ion source and $\approx 1$~mW/mm${}^2$ of cooling laser intensity.

\subsection{Ion cyclotron resonance: electronic and optical detection}
As an addition to LIF detection, SpecTrap can also perform FT-ICR (Fourier Transform Ion Cyclotron Resonance) measurements, a well-established technique for non-destructive mass and charge state spectrometry in ions traps \cite{Bla06}. By a combination of both, it is possible to gauge the electronic signal height obtained in FT-ICR to the number of observed ions as measured by LIF. Hence, a stored ion cloud can be characterized by the ion number and temperature. For FT-ICR, the ion motion is excited by a fix-phase burst such that subsequent signal pickup of the ions' oscillatory motions is efficient. A transient of that signal is recorded and its Fourier transform represents a spectrum of the mass-to-charge spectrum of ions present in the trap. 

In the present case, the signal for ion excitation is generated by an Agilent 33250A frequency generator and processed by the SpecTrap rotating wall drive \cite{Bha12}, which splits the input signal into two with a 180${}^{\circ}$ phase difference. These two signals are transmitted to two opposing ring segments, while the remaining two segments are kept at DC potential. The signal induced in the trap electrodes by the excited ions is amplified by a cryogenic amplifier mounted next to the trap and processed by a HP3589A spectrum analyser.

\begin{figure}[ht]
\begin{center}
%\begin{minipage}[c]{0.57\textwidth}
	\includegraphics[width=0.42\textwidth]{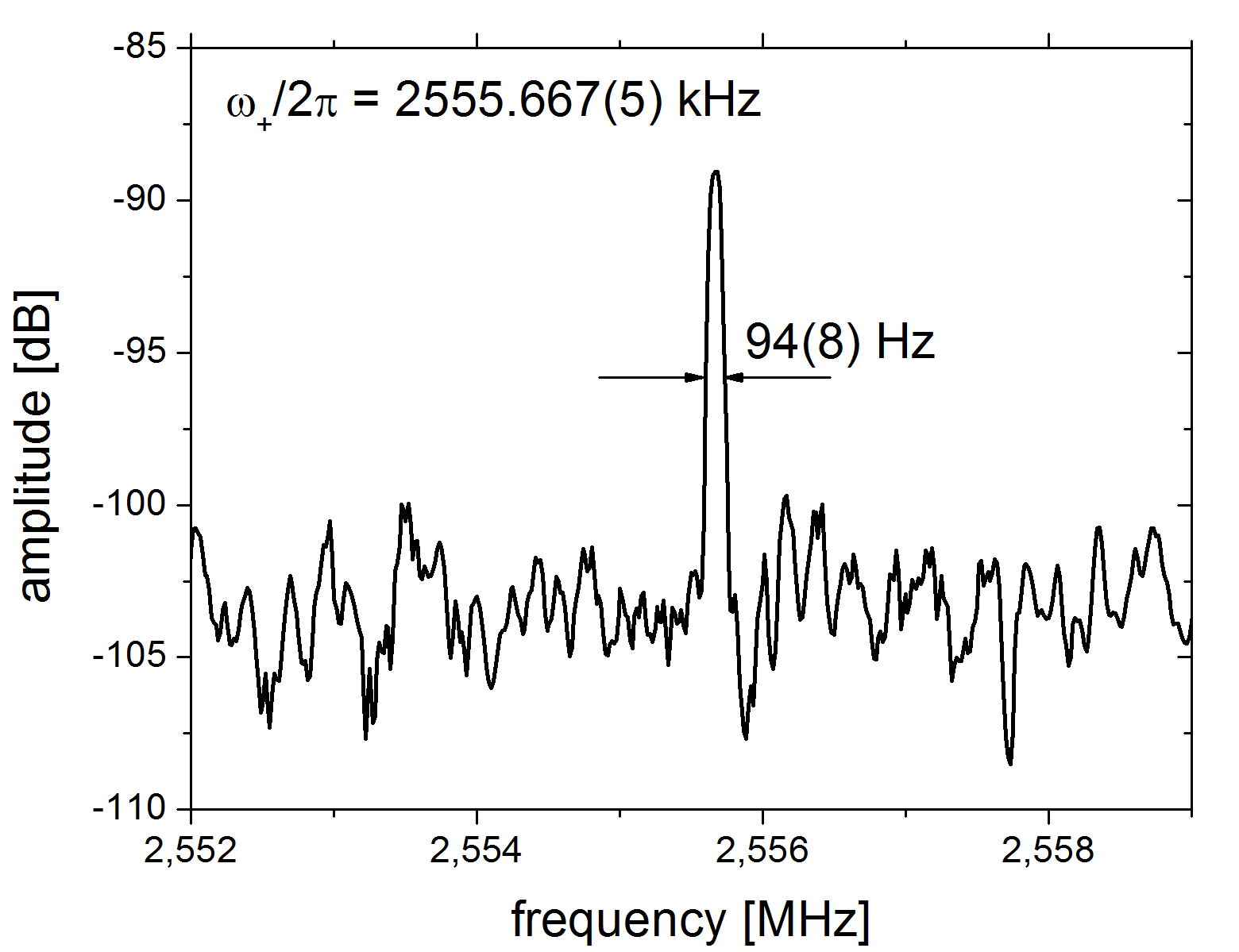}
%\end{minipage}
%\begin{minipage}[c]{0.4\textwidth}
	\caption{\small Amplitude of the FT-ICR signal of around 2000 trapped $\rm Mg^+$; $\omega_+/2\pi$ is the modified cyclotron frequency.}
	\label{fig:cycresonance}
%\end{minipage}
\end{center}
\end{figure}

The dipole excitation was performed by applying 5000 cycles of a 2.555~MHz signal, where the modified cyclotron resonance $\omega_+/2\pi$ was expected for a magnetic field of 4~T. The amplitude was set to $400\ \rm{mV}_{pp}$. Because of the short coherence time the spectrum analyser was triggered by the last excitation cycle and averaged over 10 excitation-detection rounds. The observed resonance signal is depicted in Fig.~\ref{fig:cycresonance}. The width of the resonance is $\approx$~100~Hz, while the modified cyclotron frequency can be determined with an accuracy of a few Hz. The resulting mass resolving power $m/\Delta m$ is of the order of $10^4$, while the magnetic field can be determined with a relative accuracy of $10^{-6}$. Both of these values exceed the requirements of the experiment and show that electronic and optical ion detection can be performed simultaneously.

\begin{figure}[ht]
\begin{center}
	\includegraphics[width=0.42\textwidth]{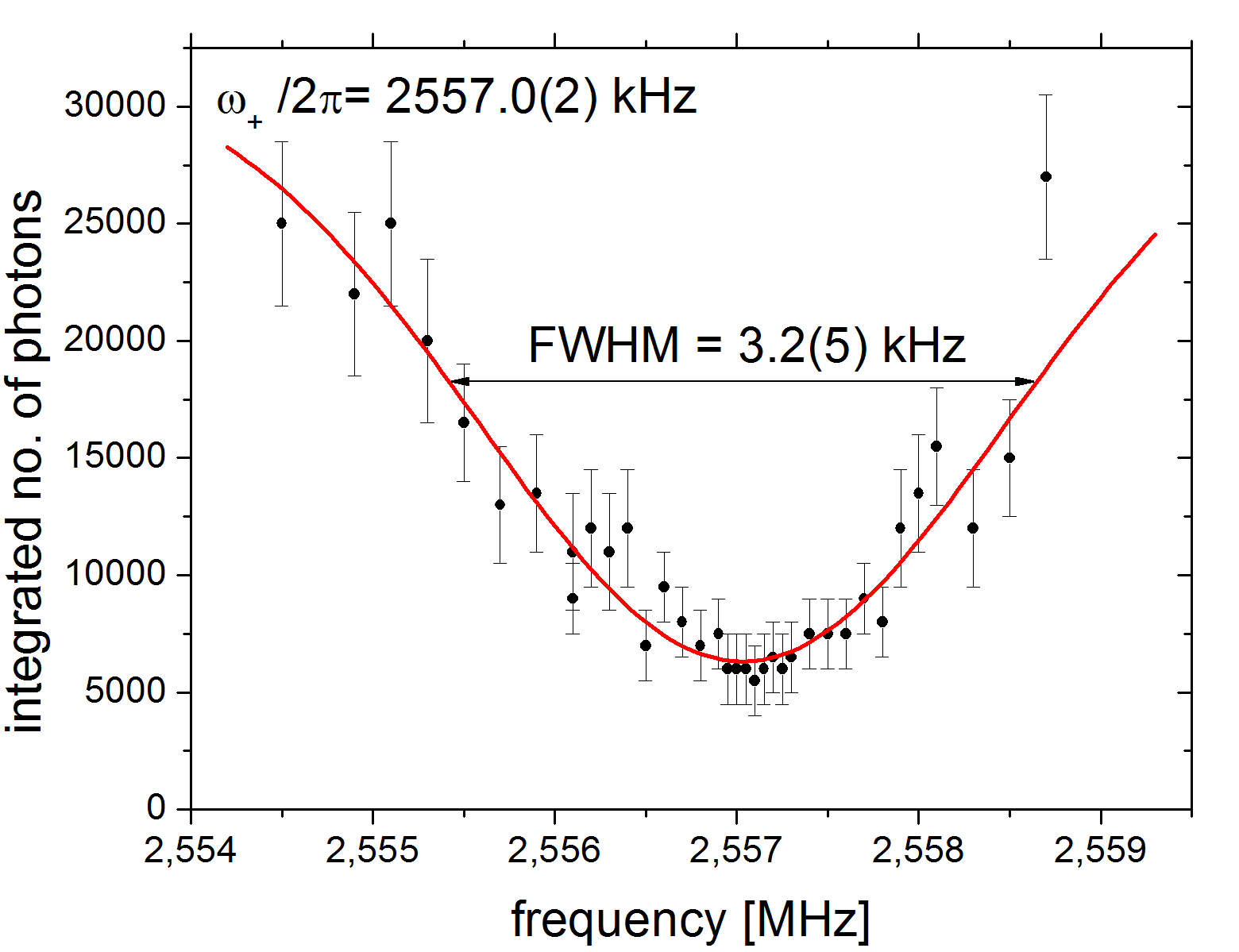}
	\caption{\small (Colour online) Ion cyclotron resonance measurement using laser-induced fluorescence. A Gauss fit of the recorded data points reveals the modified cyclotron frequency $\omega_+/2\pi$.}
	\label{fig:cyclaser}
\end{center}
\end{figure}

The transfer of energy into the cyclotron motion during excitation pushes the ions into larger orbits, where they either have a smaller overlap with the laser beam, or are even lost from the trap. This was used to perform a measurement of the modified cyclotron frequency via LIF. The ions were laser-cooled and their fluorescence recorded while applying the dipole excitation to the ring electrode. The excitation frequency was changed stepwise across the expected cyclotron resonance, while the trap was reloaded under identical conditions for each point. This resulted in a fluorescence dip seen in Fig.~\ref{fig:cyclaser}, fitted well with a Gaussian function, with the central frequency marking the resonance. It was noticed that the range of possible excitation amplitudes was rather narrow - excitation with more than 400~mV$\rm {}_{pp}$ resulted in a total loss of fluorescence or even ion loss, i.e.\ the fluorescence did not return after switching off the excitation. Conversely, amplitudes smaller than 100~mV$\rm {}_{pp}$ had little or no observable influence on the ion fluorescence.

It can be seen that the central frequencies from the two measurements, shown in Fig.~\ref{fig:cycresonance} and Fig.~\ref{fig:cyclaser}, differ by 1.33~kHz. This exceeds the statistical fitting uncertainty and was found in several repeated measurements. The systematic shift corresponds to a magnetic field difference of 2~mT. The discrepancy is ascribed to the non-ideal magnetic field and the different spatial positions where the two measurements were performed: while the FT-ICR induces a signal directly in the trap electrodes, the LIF-signal depends on the optical axis of the detector system, which is not necessarily aligned exactly along the trap radial axis. Adding a camera to the system will allow us to measure position, shape and radial extent of the cloud, which is of special interest when the rotating wall is applied. Nevertheless, it was demonstrated that electronic and laser induced fluorescence ion detection methods can be used simultaneously, with reasonably good agreement.

\subsection{Ion temperature and evidence of crystallization}
By determining the transition linewidth $\Delta\nu$ and assuming the absence of line-broadening mechanisms other than Doppler broadening, the upper limit to the ion temperature $T$ can be calculated according to \cite{foot}
\begin{equation}
	T = \frac{\Delta\nu^2 mc^2}{8\nu_0^2k_B\cdot\textrm{ln}\,2}.
\label{eq:ionfinaltemp}
\end{equation}

\begin{figure}[ht]
\begin{center}
	\includegraphics[width=0.42\textwidth]{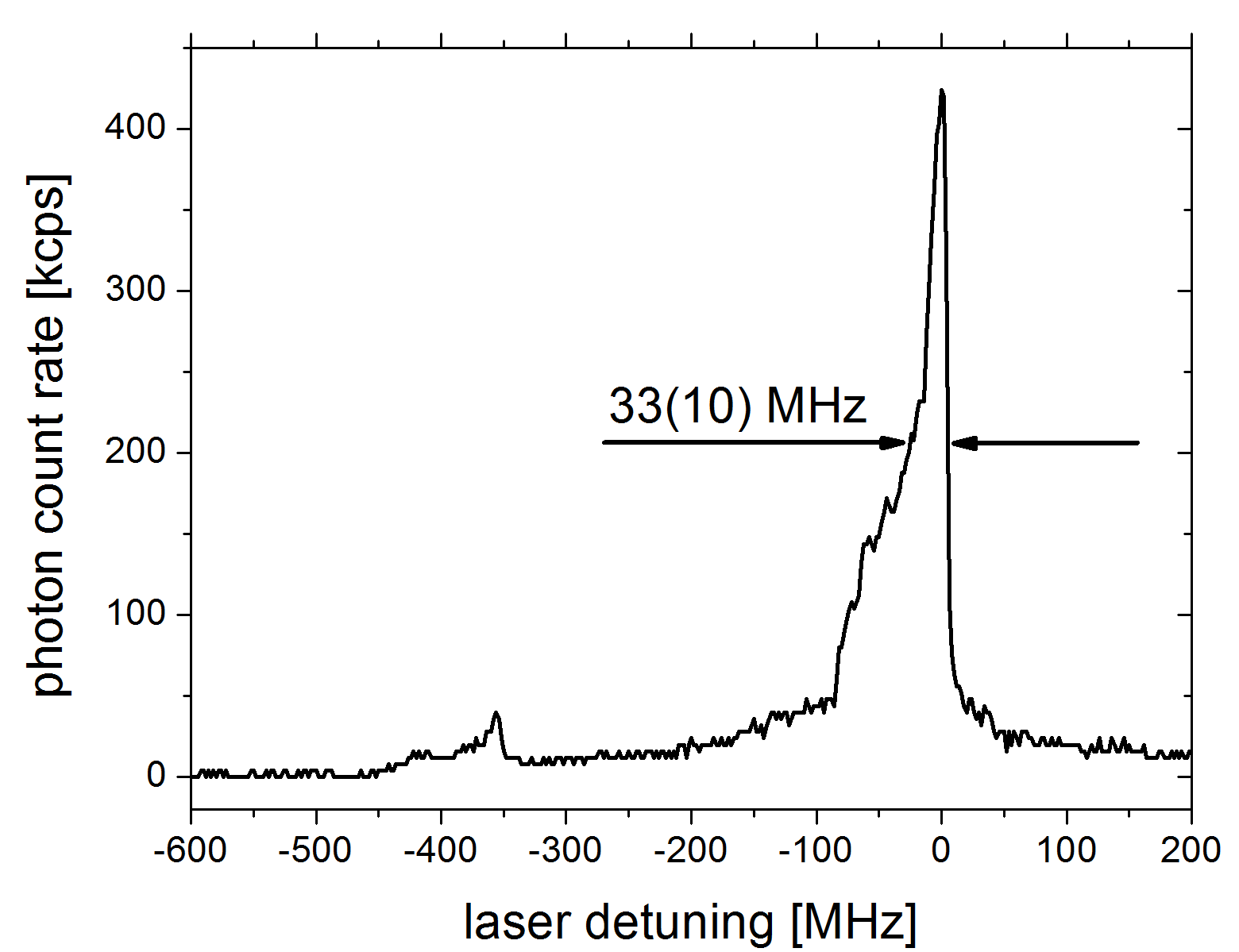}
	\caption{\small Laser induced fluorescence of trapped and laser-cooled Mg${}^+$ as a function of frequency detuning. The width of the measured transition can be used to set the upper limit to the achieved ion temperature. A precooling peak and an abrupt drop after crossing the zero is typically associated with crystalline structure of the ion cloud.}
	\label{fig:crystal}
\end{center}
\end{figure}

A series of measurements was performed in order to determine the transition linewidth of the laser-cooled $\rm Mg^+$. The laser frequency was kept 1 GHz red-detuned during ion accumulation, as well as for another 10 seconds after closing the trap. It was observed that due to the large initial ion energy this pre-cooling time was necessary for efficient laser cooling. After pre-cooling, the laser frequency was scanned over the central transition frequency of Mg${}^{+}$ and the fluorescence recorded. A typical result is shown in Fig.~\ref{fig:crystal}, where the recorded fluorescence rate was plotted against the laser frequency detuning.

After crossing the resonance frequency, ion cooling turns into heating and the fluorescence drops quickly to zero. It can therefore be safely assumed that the total FWHM of the Voigt profile is less than twice the observed width of 33(10) MHz indicated in Fig.~\ref{fig:crystal}. This value is of the same order of magnitude as the natural linewidth of the transition (42 MHz), and a deconvolution of the Doppler and the natural linewidth contribution to the line profile needs to be performed \cite{Oli77}. The deconvoluted value for the Doppler width can be inserted into Eq.~\eqref{eq:ionfinaltemp} and an upper limit for the ion temperature is obtained
\begin{equation}
	\Delta\nu_D \leq 39(11) \ \rm{MHz}, \qquad T \leq 0.06(3) \ \rm K.
\label{eq:finaltemp}
\end{equation}

By experience from a similar experiment \cite{Gru05}, the typical volume of such an ion cloud is of the order of 0.5 mm${}^{3}$, resulting in an ion number density of around $n=4000$ ions/mm${}^{3}$ for $\rm Mg^+$ ions stored in SpecTrap. Under such conditions the single-particle description begins to break down and the ion cloud has to be treated as a non-neutral plasma. The plasma coupling parameter, describing the ion Coulomb coupling intensity in one-component plasmas, is given by \cite{Gil88}
\begin{equation}
	\Gamma_p = \frac{q^2}{4\pi\epsilon_0 a_s k_BT}; \qquad
	a_s = \sqrt[3]{\frac{3}{4\pi n}}
\label{eq:plasmacoupling}
\end{equation}
where $a_s$ is the Wigner-Seitz radius. Gilbert and co-workers have predicted that for coupling parameters $\Gamma_p\geq2$ the plasma starts gradually to exhibit liquid-like properties \cite{Gil88}. According to Eq.~\eqref{eq:plasmacoupling}, for $\rm Mg^+$ trapped and cooled in SpecTrap ($n\approx4000$ ions/mm${}^{3}$ and $T\approx 60 \ \rm mK$) the plasma coupling parameter amounts to $\Gamma_p\geq 7(4)$, such that strong ion coupling may be assumed. After the planned introduction of HCI into the trap, the temperatures of the two components are expected to roughly equalize, resulting in a much larger $\Gamma_p\geq 1000$ for high charge states \cite{Gru01}.

Studies have already shown that for sufficiently low temperatures, a trapped ion cloud exhibits a structural change and its spectrum resembles the one of single ions \cite{Die87,Gud95,Oka09}. A similar behaviour was observed for the trapped $\rm Mg^+$ in SpecTrap and is shown in Fig.~\ref{fig:crystal}. A small pre-cooling peak appears at the point where the transition's Doppler broadened half-width becomes smaller than the laser detuning (here at $\approx$~400~MHz), after which the fluorescence disappears and can be observed again only close to the natural linewidth of the transition. Such structures were observed also in \cite{Die87,Gud95,Oka09} and mark the transition of the stored ion plasma from a non-correlated to a strongly coupled state. Because of strong cooling and simultaneous reduction of the Doppler width, the fluorescence close to the resonant frequency is characterized by a sharp asymmetric shape, followed by an abrupt drop to zero after crossing the central frequency.

The spectrum shown in Fig.~\ref{fig:crystal} was recorded with 1.1~mW/mm${}^2$ of laser power and a 100~MHz/s frequency sweep. It has been observed that different pre-cooling times cause different positions of the pre-cooling peak with respect to the main one, moving them closer together for shorter pre-cooling times. This structure was, however, not observed for very short pre-cooling times below roughly 8 s, which were also typically followed by much smaller or no detectable fluorescence. However, after allowing sufficient pre-cooling time and observing the crystalline structure, a smaller, sharp fluorescence peak was observed at resonance even when scanning the laser frequency in the opposite direction.

\section{Summary and Conclusion}
We have performed systematic measurements with laser-cooled ${}^{24}$Mg${}^+$ ions stored in a Penning trap. These ions were externally produced, transported, captured and stored in the trap for subsequent measurements. Using both optical and electronic non-destructive detection techniques, the properties of stored ion clouds were determined. Combining electronic with optical detection, it is possible to determine stored ion numbers down to the single ion level and to characterize the stored ion cloud with respect to its temperature, storage time and related properties. Laser cooling was achieved to temperatures below 0.1~K and evidence of ion crystallization was found. Such laser-cooled ions are ideal for sympathetic cooling of simultaneously trapped ion species which lack a suitable level scheme, as for example the highly charged ions that will be available in the near future. All necessary prerequisites for precision spectroscopy of externally produced highly charged ions to be delivered by the HITRAP facility at GSI were demonstrated in the present setup with singly charged magnesium ions. Further investigation of sympathetic cooling of mid-Z highly charged ions in SpecTrap will be carried out in the near future. Finally, the measurements of forbidden transitions in heavy highly charged ions will open the way to precision tests of QED calculations in extreme fields and to the determination of fundamental constants.

\section{Acknowledgements}
This work is supported by HGF under contract number VH-NG-148, BMBF under contract numbers 06MS7191, No. 05P12RDFA4 and 06DA9020I as well as EPSRC under the grant number EP/D068509/1. We also thank LLNL and the former RETRAP group for their support and loan of the equipment to GSI.


\end{document}